# Modernising the Design and Analysis of Prevalence Surveys for Neglected Tropical Diseases

Peter J Diggle*[1]  Claudio Fronterre[1] Katherine Gass[2]  Lee Hundley[2]
Reza Niles-Robin[3]  Annastacia Sampson [3] Ana Morice[4] Ronaldo Carvalho Scholte[4]

1. CHICAS,Lancaster Medical School, Lancaster University, Lancaster UK
2. Task Gorce for Global Health, Decatur, Georgia, USA
3. *Neglected Tropical Disease Program, Vector Control Services, Ministry of Health, Georgetown, Guyana*
4. Neglected, Tropical, and Vector-Borne DiseasesUnit, Communicable Diseases and Environmental Determinants of Health Department, Pan American Health Organization, Washington, D.C., USA

## Summary

Current WHO guidelines set prevalence thresholds below which a Neglected Tropical Disease can be considered to have been eliminated as a public health problem, and specify how surveys to assess whether elimination has been achieved should be designed and analysed, based on classical survey sampling methods. In this paper we describe an alternative approach based on geospatial statistical modelling. We first show the gains in efficiency that can be obtained by exploiting any spatial correlation in the underlying prevalence surface. We then suggest that the current guidelines' implicit use of a significance testing argument is not appropriate; instead, we argue for a predictive inferential framework, leading to design criteria based on controlling the rates at which areas whose true prevalence lies above and below the elimination threshold are incorrectly classified.  We describe how this approach naturally accommodates context-specific information in the form of georeferenced covariates that have been shown to be predictive of disease prevalence. Finally, we give a progress report of an ongoing collaboration with the Guyana Ministry of Health Neglected tropical Disease program on the design of an IDA (Ivermectin, Diethylcarbamazine and Albendazole) Impact Survey (IIS) of lymphatic filariasis to be conducted in Guyana in early 2023.

## Introduction

The WHO Roadmap (WHO, 2020) sets out disease specific, as well as cross-cutting, targets to meet the eradication, elimination and control goals for the 20 neglected tropical diseases (NTDs). An important emphasis of this new Roadmap is the focus on measures of impact.  For the preventive chemotherapy treated NTDs, impact is measured using prevalence surveys designed to assess whether the prevalence of a disease indicator is below a threshold that defines elimination of transmission or elimination as a public health problem. Typically, elimination will result in the reduction or cessation of mass drug administration (MDA). Consequently, it is important that these prevalence surveys be sufficiently rigorous to instil confidence that elimination has indeed been achieved, while adhering to a design that is not resource prohibitive for ministries of health and disease-specific programs to implement.

---

*Author for correspondence (p.diggle@lancaster.ac.uk).
†Present address: CHICAS,, Lancaster Medical School, Lancaster University, Lancaster LA1 4YF, UK.

# Current design and analysis guidelines

Each NTD has its own set of guidelines for the design and analysis of a survey to establish whether or not a particular area, termed an *evaluation unit* (EU), has or has not achieved elimination. In this paper, we focus on the case of lymphatic filariasis (LF), for which the Roadmap target is 'elimination as a public health problem' and the relevant guideline is the Transmission Assessment Survey, or TAS (WHO, 2011). The key purpose of the TAS in areas where *Anopheles* or *Culex* is the principal vector, is to determine whether the mean antigenaemia prevalence is below 2% (WHO, 2011, Section 7).

The LF TAS is a 30-site survey, with sites typically selected by drawing up a list of all potential sites and taking a systematic sample of every kth site from a random starting point, where k is chosen to result in a sample of size 30. The total sample size is prescribed, based on the total target population in the EU, as is a corresponding critical cut-off of positive cases for measuring the target threshold with a predetermined level of sampling error. Under the current guidelines, the decision to declare an EU as having eliminated LF as a public health problem is based on the results of three TAS surveys, conducted independently at two-year intervals, each indicating that the number of positive cases is below the critical cut-off value.

A simple example, a variation on one previously given in Diggle et al (2021), illustrates a key limitation of this approach. The left-hand panel of Figure 1 shows a hypothetical set of prevalence data taken from locations indicated by open circles, with radius proportional to the observed prevalence at the location in question. In the right-hand panel, the locations are the same, but the prevalence values have been randomly permuted. Even without conducting any formal statistical analysis, the left-hand panel shows clear spatial structure which might, for example, present the reader with an opportunity to bet that if additional surveys were conducted at locations near the upper central and lower left parts of the square, they would likely yield low and high observed prevalence, respectively. By construction, the centre panel presents no such opportunities, butut the TAS would treat both sets of data as identical. Put another way, in the left-hand panel the value of the observed prevalence at a particular location conveys partial information about the underlying true prevalence at, *and in the neighbourhood of*, that location, and by exploiting this we can do better than the current TAS guideline. For completeness, the right-hand panel of Figure 1 shows the true prevalence surface from which the data in the left-hand panel were generated.

# The geostatistical alternative

The example summarised in Figure 1 may convince the reader, as it does the authors, that *location matters*. Expressed in more formal statistical terms, the fact that observed values of prevalence are predictive of unobserved values at nearby locations is a consequence of the existence of *spatial correlation*. If P1 and P2 are the (unknown) values of prevalence at two locations, $\sigma^2$ is the *a priori* variance of P1 or P2 and $\varrho$ is the correlation between P1 and P2, then the variance of P2 given P1 is $\sigma^2(1-\varrho^2)$, which is less than $\sigma^2$ unless $\varrho=0$. The correlation is *spatial* if the value of $\varrho$ depends on the corresponding locations, typically as a function of the distance between them. Model-based geostatistical methods (Diggle, Moyeed and Tawn, 1998; Diggle and Ribeiro, 2007; Diggle and Giorgi 2019*)* explicitly recognise the possibility of spatial correlation in prevalence, but do not impose it; rather, they estimate the nature and strength of spatial correlation from the available data as an integral part of the overall analysis protocol.

The simplest form of prevalence data is a set of triplets *($x_i$, $n_i$, $y_i$): i=1,…,m*, where the $x_i$ denote locations within the study-area, $n_i$ the number of individuals sampled and tested at location $x_i$, and $y_i$ the corresponding number of positive test results. The basic geostatistical model for prevalence data can then be specified as follows:

> M1: conditional on an unobserved true prevalence surface *P(x),* the $y_i$ are independent realisations of binomially distributed random variables with denominators $n_i$ and probabilities *P($x_i$)* of a positive test result;
> 
> M2*:* log *P(x)/(1- P(x)) = S(x)* is an unobserved, stationary Gaussian stochastic process with mean µ, variance $\sigma^2$ and correlation function $\varrho(u)$, where *u* denotes distance.

In M2 above, the mean and variance have the standard, elementary definitions; in particular any stationary Gaussian process can be standardised by subtracting the mean and dividing by the standard deviation to give a process $S^*(x) = (S(x) – \mu)/ \sigma$ with mean zero and variance 1. A Gaussian process is one whose values at any finite set of locations follow a multivariate Normal distribution. It follows that the properties of the standardised process are completely determined by its correlation function, $\varrho(u)$. The most important property of $\varrho(u)$ is its range, conventionally defined as the distance $u$ at which $\varrho(u)=0.05$. Additionally, the shape of $\varrho(u)$ can affect the analytic smoothness of the prevalence surface; for example, the widely used Matérn class of correlation functions (Matérn, 1960) includes a parameter that controls the degree of mean-square differentiability of $S^*(x)$.

Figure 2 illustrates how the form of the correlation function manifests itself in the "patchiness" of the corresponding prevalence surface. The three panels in the upper row of Figure 2 show surfaces on a square with side-length one; the corresponding panels in the lower row show their respective correlation functions. The left-hand and centre panels were simulated from processes with correlation ranges 0.15 and 0.3, respectively; the larger value of the correlation range leads to larger patches of high and low values of the simulated surface. The centre and right-hand panels were simulated from processes that both have correlation range 0.3 but are, respectively, continuous but non-differentiable (centre panel), and differentiable (right-hand panel). The differences between these two are more subtle than in the previous comparison; the "patches" are similar in size in both, but the gradations from high to low are visually smoother in the right-hand panel. This suggests, correctly (Zhang, 2004), that it would be difficult to distinguish between the two on the basis of prevalence data alone, which are necessarily spatially discrete.

Under the model defined by M1 and M2 above, the application of general statistical principles, specifically likelihood-based and predictive inference, delivers the probability distribution of the complete surface $P(x)$ over the study-region, conditional on the prevalence data. By drawing samples from this *predictive probability distribution*, the user can evaluate the predictive probability distribution of whatever attribute, or set of attributes, of $P(x)$ is required for the practical problem at hand. In practice, we use Markov chain Monte Carlo methods (Gilks et al, 1996) to sample the values of $P(x)$ at a set of points $x_j$ on a finely spaced grid that covers the study-region.

In the current context, the question of primary interest is: has a particular EU achieved elimination? The model-based geostatistical answer proceeds as follows:
- P1: the predictive target is the population-weighted average of $P(x)$ over the EU in question,
$$T = \int pd(x) P(x) dx / \int pd(x) dx \qquad (1)$$
where $pd(x)$ is the population density at $x$;
- P2: model-based geostatistics delivers the predictive probability, $q$ say, that $T<c$, where c is the WHO-declared threshold to define elimination (2% or 1%, depending on the locally predominant vector species);
- P3: elimination is declared if $q$ exceeds a pre-agreed value, say 0.95 or 0.99.

In this paradigm, we are using the word prediction in the following precise sense: a *prediction* is a probability statement about an unobserved random variable conditional on all available data.

## Implications for study-design

Spatial correlation also has implications for efficient study-design; specifically, *how many* sampling locations are needed and *where* within the study-area should they be placed?

With regard to the *where*, it has been known at least since 1960, through the pioneering work of Bertil Matérn (Matérn, 1960) that in the presence of spatial correlation, if the objective is to estimate the spatial average of a phenomenon over a designated area, then taking measurements of the phenomenon at a completely random sample of locations is inefficient. This is not to deny the importance of using a probability-based sampling algorithm to remove the risk of subjective bias, but for efficiency the algorithm should ensure that sampled locations are spatially regulated, meaning that the collection of sample locations covers the study-area more evenly than would be expected under completely random sampling (Chipeta *et al*, 2017). This does presume that any location within the study-region is a potential sampling location. In practice, sampling can only take place among discrete communities or schools and, depending on their spatial distribution, this may lessen or entirely lose the benefits of spatially regulated over random sampling. In fact, the current guideline recommends an informal version of spatially regulated sampling, suggesting that the list of potential sampling locations "should be numbered by geographical proximity, as opposed to alphabetical order" (WHO, 2011, Annex 5).

Whether spatially regulated or random, it is usually worth considering a stratified sampling scheme whereby the study-region is partitioned into pre-specified sub-regions according to the aims of the study. In the Guyana LF case-study that we present later in the paper, each evaluation unit was designated as a stratum and we used a mildly regulated sampling scheme within each stratum.

With regard to the *how many*, the current WHO guidelines use, if only implicitly, a significance testing paradigm. Consider the following extract for settings where the target threshold for elimination is <2% antigenaemia (WHO, 2011, Section 7.3.7):
"The sample sizes and critical cut-off values were chosen so that an EU has:
1) at least a 75% chance of passing if the true prevalence of antigenaemia is 1% (half the target level); and
2) no more than about a 5% chance of passing (incorrectly) if the true prevalence of antigenaemia is ≥2%"

Here, "passing" the TAS means that the total number of positive test results found within the EU in question is less than or equal to a prescribed number (the "critical cut-off value"). In other words, the 75% and 5% chances are probability statements about a statistic calculated from the data given that the underlying EU-wide antigenaemia prevalences are 1% and 2%, respectively. Or, expressed more formally, "passing" the TAS is equivalent to rejection, at the 5% level of significance, of the hypothesis that EU-wide prevalence is 2%, the test being designed to achieve a power of 75% when the true EU-wide prevalence is 1%. However, the operational significance of passing or failing the TAS is that MDA will likely cease or continue, respectively; if the number of positive cases "is greater than the critical cut-off number, MDA should continue in the EU for two more rounds" (WHO, 2011, section 7.3.8). But no incomplete sample can determine the true EU-wide prevalence – the best it can do is indicate *how likely* it is that the true prevalence is or is not "equal to or greater than 2%." This requires a probability statement about the true prevalence given the data, not the other way around, i.e. a *predictive inference*, which is exactly what the model-based geostatistical approach delivers, at P2 above. In the authors' view, the correct re-expression of the TAS design requirements should be in terms of its positive and negative predictive values. Under this predictive inferential paradigm, an EU would pass the TAS if and only if *the predictive probability that the EU-wide population-weighted prevalence is greater than 2% is at most 0.05 (5%)* and the design requirement would be that among EUs that pass the TAS by this criterion, *the probability that their EU-wide population-weighted prevalence is in fact less than 2% should be at least 0.75 (75%)*.

The "how many" question has two dimensions: how many sampling locations (schools/communities); and how many individuals per sampling location? Whatever the analysis approach, the selection of individuals *within* each sampling location should be randomised to give every member of the population an equal chance of being sampled. However, even when a random or "as if random" sampling method can be adopted within each sampling unit, complete compliance never happens in practice. For this reason, a minimal safeguard against selection bias is to record the number (and, ideally, key characteristics such as age and gender) of non-compliant individuals to enable a partial check for possible selection bias at the analysis stage.

Within the model-based geostatistical approach, there is no requirement for every potential sampling location to have an equal chance of being included. The requirement for validity of the geostatistical inference is that the sampled locations, are stochastically independent of the underlying prevalence surface. For example, this condition would be met by applying a probability sampling scheme within each of a number of pre-defined strata, but would be violated if field-teams were instructed to choose sampling locations that seemed to them to be particularly favourable for disease transmission, so-called preferential sampling (Diggle, Menezes and Su, 2010).

Efficiency is harder to quantify, because the variance of our predictive target *T*, as defined at equation (1), is determined by a complex combination of the spatial variation in the population density, *pd(x)*, over the EU in question, the spatial covariance structure of the underlying prevalence surface, *P(x)*, and the binomial variation in the location-specific prevalence data conditional on *P(x)*. In view of this, a simple rule of thumb that places all EUs on an equal footing is to sample equal numbers of individuals per sampling location, and equal numbers of sampling locations per EU.

A third design consideration, which directly affects the subsequent analysis protocol, is what, if any, additional information we should collect on potential covariates. The minimal requirement for a variable to be included as a covariate is that its value is available at every point *x* in the study-region, which in practice means that it is available as a raster image at a sufficiently fine resolution that its true value does not vary materially within a single pixel. Examples that are readily available as remotely sensed images and have proved useful in specific applications include elevation or various vegetation indices (for a review, see Huang *et al*, 2021). Adding a covariate, *d(x)*, say, to the geostatistical model is straightforward – we simply replace M2 in the model specification by

   M2*: $\log P(x)/(1- P(x)) = \alpha + \beta d(x) + S(x)$, where now $S(x)$ is an unobserved, stationary Gaussian stochastic process with mean 0, variance $\sigma^2$ and correlation function $\varrho(u)$, where *u* denotes distance.

Against this, adding covariates increases the number of parameters that need to be estimated and, as presented in M2* above, makes the additional assumption of a *linear* association between the covariate and the log-odds of prevalence; for example, Diggle *et al* (2007) used splines to capture non-linearities in the associations. Finally, any covariates that measure relevant characteristics of individuals rather than locations, however relevant they may be as contributors to an individual's disease risk, can only be used if they are reduced to average values over a set of areas that partition the whole study-region, in which case they are susceptible to ecological bias (Greenland and Morgenstern, 1990) that is liable to attenuate their effects. The authors' current opinion is that, at the design stage, covariates should be introduced cautiously, if at all. At the analysis stage, covariates can be introduced if they deliver demonstrably more precise predictions.

## Case-study: designing the 2023 Guyana LF IDA Impact Survey

In this Section, we describe the design of an IDA (Ivermectin, Diethylcarbamazine and Albendazole) Impact Survey (IIS) to be conducted in Guyana early in 2023. The IIS is a prevalence survey for making stop treatment decisions when triple drug therapy is used for LF mass treatment (as opposed to the standard two-drug regimen). The design of the IIS is analogous to the TAS with the notable exception that the target threshold is <1% microfilaremia (mf), which must be measured in adult populations, hence sampling locations are restricted to villages, not schools. Guyana is partitioned into 27 ADM2 units, the second-level administrative division of the country. The IIS is required to cover 10 Evaluation Units (EUs), each of which corresponds to one or more ADM2 units that collectively span eight of the country's 10 regions. The set of potential sampling locations consists of the villages that lie within this set of 10 EUs. The goal of the IIS is to predict, for each EU, whether the EU-level prevalence of LF as diagnosed by the presence of microfilariae by microscopy (mf prevalence), is or is not less than 1%. A model-based geostatistical analysis of the IIS data delivers, for each EU, the probability that the true EU-level mf prevalence is less than 1%, and the EU *passes* the IIS if this probability is at least equal to a pre-specified value *q*. The performance of any particular study-design is measured by its positive and negative predictive values, defined as follows. The *positive predictive value* (PPV) is the probability that an EU which passes the IIS has a true mf prevalence less than 1%; the *negative predictive value* (NPV) is the probability that an EU which fails the IIS has a true mf prevalence greater than 1%. The design requirement is to choose which villages to sample in each EU, and how many people to test within each sampled village, so as to achieve the following.
   (1) PPV ≥ *0.95*
       The probability that an EU which passes the IIS has a true mf prevalence less than 1% is *at least* 0.95, i.e. *q*=0.95
   (2) NPV≥ *0.75*
       The probability that an EU which fails the IIS has a true mf prevalence greater than 1% is *at least* 0.75.

The potential sampling locations were initially taken from a national gazette with 4412 entries. The study-design consists of choosing the set of *k* villages to be sampled in each EU and the number, *m*, of individuals to be sampled per village so that the design requirements (1) and (2) above are both satisfied. Sampled villages in each EU were chosen according to a spatially regulated design in which no two sampled villages could be separated by less than 2km.

To evaluate any candidate design, we first needed to develop a working model for the unknown Guyana-wide prevalence surface. To do this, we fitted a geostatistical model, as defined by M1 and M2 above, to the most recent available prevalence data, consisting of test results from a total of 7839 individuals across 214 villages, collected between 2018 and 2021. Village-level empirical prevalence showed a highly skewed distribution, with 184 zeros, a maximum of 0.5 and a mean of 0.012 (Figure 3). For the geostatistical model, we specified a continuous, non-differentiable correlation function, $\varrho(u) = \exp(-u/\phi)$, and obtained Monte Carlo maximum likelihood estimates of the model parameters, $\mu$, $\sigma^2$ and $\phi$, with confidence intervals whose width reflects the sparseness of the data (Table 1).

To simulate current prevalence surfaces throughout the study-region we used the fitted model with a shift of the log-odds prevalence surface to correspond to a study-region-wide average prevalence 1%, the threshold for elimination. From each simulated surface we calculated the population-weighted prevalence for each EU and classified EUs as eliminated or not, accordingly. For each candidate design, we then simulated prevalence data by applying the design to the simulated prevalence surface. Next, we re-fitted the geostatistical model to the simulated data and, for each EU, calculated the predictive probability that the population-weighted prevalence exceeded 1%. If this probability was greater than 0.95, we classified the EU as *elimination indicated*, and conversely. Finally, we estimated the design's NPV as the proportion, over 1000 repeated simulations, of EUs for which elimination was not indicated, among those whose actual prevalence was greater than 1%. Table 2 shows the resulting estimates of NPV for a range of combinations of *k* and *m*, and suggests that the design requirement can be satisfied by sampling 60 individuals within each of 10 villages in each EU. However, these estimates of NPV assume 100% compliance among sampled individuals, which is never achieved in practice. To allow for this, we recommended aiming to sample *m*=100 individuals in each of *k*=10 villages in each EU.

The recommended design turned out to be impractical because the gazette included locations that, on subsequent inspection, were found to be uninhabited, and some whose populations were less than 100. We excluded the former, but retained the latter because, as indicated in Table 1, the design performance was more strongly affected by *k* than by *m*, and on the understanding that in such villages the field teams would attempt to sample all eligible individuals. Finally, to allow for the possibility that some village communities might be unwilling to take part in the study, we selected an additional two villages per EU where feasible, to be held in reserve in case needed.

## Discussion

The recommended design represents a substantial saving over the current TAS guidelines, which require 30 villages to be sampled in each EU. We have found similar gains in efficiency in other applications, including LF in Ghana (Fronterre et al, 2020) and soil-transmitted helminths in Kenya, Sierra Leone, and Zimbabwe (Johnson et al, 2021).

The size of the gain in efficiency of the geostatistical approach by comparison with current guidelines depends primarily on the range of the spatial correlation of the underlying prevalence surface; where there is little or no evidence of spatial correlation, the geostatistical approach confers little or no benefit. A secondary consideration is the number of EUs into which the study-area is partitioned. In different parts of the world, EUs can vary in size by orders of magnitude yet, to take the TAS as an example, the current guideline specifies a 30-site survey per EU irrespective of its size. Roughly speaking, 30 or so is about the minimum number of locations needed to fit a geostatistical model without strong prior assumptions about the model parameters. It follows that for a single-EU survey the geostatistical approach delivers little or no reduction in the required sample size; but it can still deliver more precise predictions of the EU-wide prevalence (cf Figure 1) and, potentially more importantly, can identify within-EU variations in local prevalence that could inform local public health action

Our design calculations are based on an assumed model fitted to historic data and, like any sample size calculation, their accuracy rests on the adequacy of the fitted model as a representation of the current state of the disease process. In the Guyana LF example, the historic data were only two or four years old, and it is reasonable to assume that the spatial distribution of disease would not have changed radically in the intervening years. For this reason, the only adjustment we made to the fitted model was a small shift in the intercept parameter μ to give a country-wide average prevalence of 0.01, the threshold for elimination of the disease as a pubic health problem. When the gap between the most recent available data and the present is longer, more sophisticated ways to simulate the current state of the disease process, for example to reflect different levels of MDA since the data were collected, or by using a mechanistic transmission model (Retkute et al, 2021; Toulopou et al, 2022).

As prevalence declines, fitting a geostatistical model becomes problematic. In the Guyana LF example, the many instances of no positive detections resulted in very imprecise model parameter estimates. The actual data that will be collected to determine the elimination status of each EU are likely to be similar, in which case we propose to allow for parameter uncertainty by adopting a Bayesian analysis with priors based on the model fitted to the historic data.

From the perspective of programmatic resources, the advantages of the geostatistical approach are clear.  The current IIS design would have required visiting 30 clusters in each EU and  sampling 100 individuals per cluster where possible; in some villages the number of eligible individuals is less than 100.  For the 10 EUs in Guyana, this would have amounted to visiting 300 villages and sampling 26,355 individuals; a massive undertaking for any public health program.  The geostatistical model enabled the Guyanese programme to reduce the number of clusters, resulting in a total sample size of 100 villages and up to 10,000 individuals - at least a 62% reduction.  In Guyana, and indeed in most prevalence surveys globally, the greatest survey expense is the cost and time required for the teams to visit each cluster. By dramatically reducing the total number of clusters to be visited, the geostatistical model resulted in a major saving of programme resources, which can then be redirected towards other priority activities.

While the benefits of using a geostatistical model to design prevalence surveys are apparent in terms of cost and efficiency gains, some practical considerations are currently limiting their widespread uptake for NTDs. Firstly, this approach requires historical prevalence data to develop the model. For many NTD programs baseline data may be sparse and lack georeferencing. This latter issue can be overcome by approximating the locations on a map; however, staff turnover and poor record keeping have created some instances where the cluster level information is lost entirely. Fortunately, the growing appreciation for the advantages of geostatistical approaches and ubiquity of cell phones have led to an increase in georeferenced data capture for NTDs, paving the way for more widespread use of geostatistical methods in the future.  Another important limitation is the complexity of the methods themselves. At present, to take advantage of these tools requires the time and expertise of a well-trained statistician; a position seldom found among the slim NTD programme staff. Furthermore, Ministries of Health are increasingly reluctant to have national data shared with outside parties.  The solution to this limitation is two-fold: creation of user-friendly open-source geostatistical software tools that bring geostatistical methods into the hands of national NTD programmes; and more in-country statisticians trained at least to MSc-level who can design and analyse geospatial surveys with little or no external supervision.

Prevalence surveys are the backbone of NTD programs' monitoring and evaluation frameworks. As more countries achieve success through MDA, there is a growing need to conduct prevalence surveys to determine whether mass treatment can be stopped. In this paper we have demonstrated how prevalence surveys using a geostatistical design can lead to gains in efficiency, compared to the status quo survey designs, by exploiting the spatial correlation in the underlying prevalence surface.  For NTD programmes and their stakeholders, the ability to leverage existing programme data to achieve major reductions in cost and resources without sacrificing predictive precision represents an important opportunity to modernise NTD surveys. We hope to see greater global uptake of geostatistical prevalence surveys to measure the impact of NTD programs as they work towards the 2030 goals.

# Acknowledgments

We are grateful for the support profived by the PAHO Guyana team: Dr. Rainier Escalada and Mrs Nathely Mars-Caleb

# Tables

Table 1. Monte Carlo maximum likelihood parameter estimates and 95% confidence intervals for the model fitted to LF prevalence data from 214 villages in Guyana. The asymmetric confidence intervals for $\sigma^2$ and $\phi$ arise because these parameters were optimised on the log-transformed scale to improve the Normal approximation to the sampling distribution of the Monte Carlo maximum likelihood estimates.

| Parameter | Estimate | 95% confidence interval |
|---|---|---|
| $\mu$ | -7.16 | (-9.19, -5.13) |
| $\sigma^2$ | 4.45 | (1.24, 16.00) |
| $\phi$ | 0.46 | (0.11, 1.89) |

Table 2. Estimated values of NPV for designs with *k* sampled villages per EU and *m* sampled individuals per sampled village. Each entry in the table is an estimate from 1000 replicate simulations.

|  | *m*=60 | *m*=80 | *m*=100 |
|---|---|---|---|
| *k*=5 | 0.554 | 0.620 | 0.620 |
| *k*=10 | 0.768 | 0.769 | 0.815 |
| *k*=15 | 0.827 | 0.852 | 0.876 |

# Figures

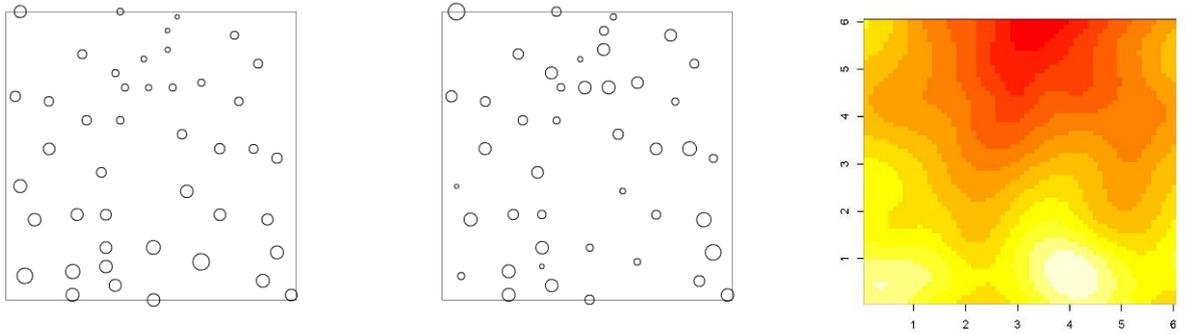

**Figure 1.** Left-hand panel: hypothetical values of prevalence at a set of locations, each indicated by an open circle with radius proportional to prevalence at the location in question. Centre panel: the same values of prevalence randomly permuted over the locations. Right-hand panel: heat map (red=low through yellow to white=high) of the prevalence surface used to generate the data shown in the left-hand panel. Adapted from Diggle et al (2021).

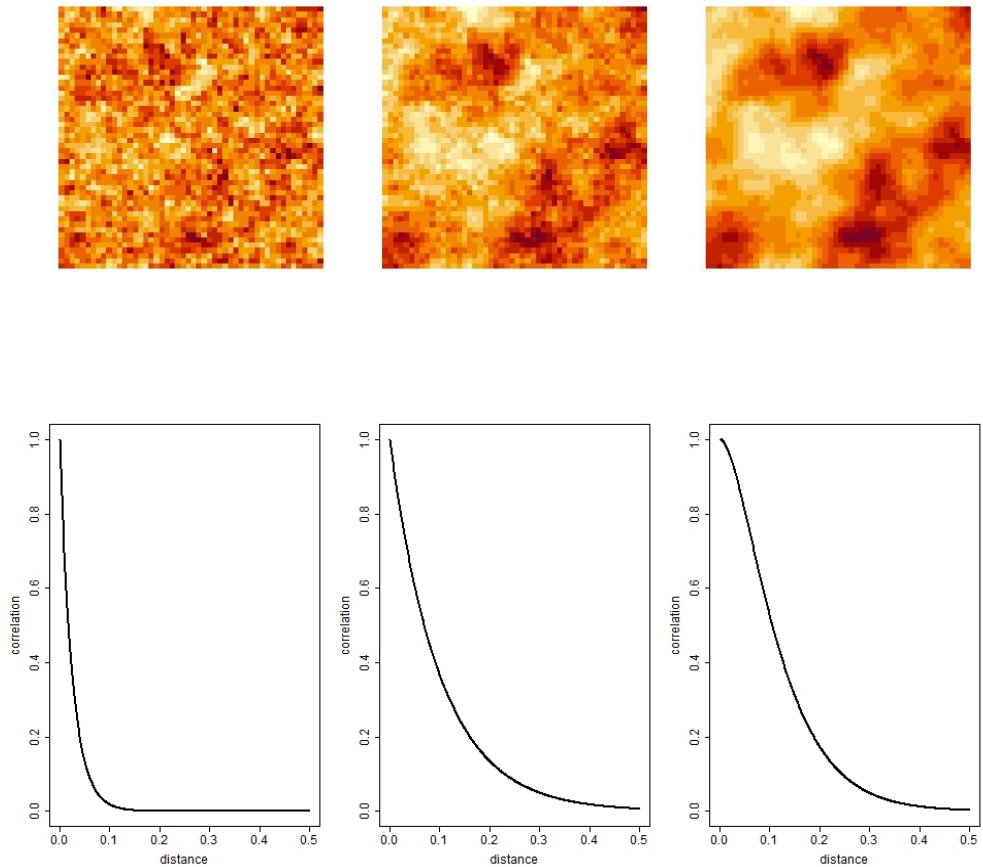

Figure 2. Some realisations (upper row) of three standardised Gaussian processes $S^*(x)$ on a square of side-length one, and their correlation functions (lower row). In the left-hand and centre panels, the processes are continuous but non-differentiable with correlation functions whose ranges are 0.15 and 0.3, respectively, In the right-hand panel, the process is differentiable with correlation range 0.3.

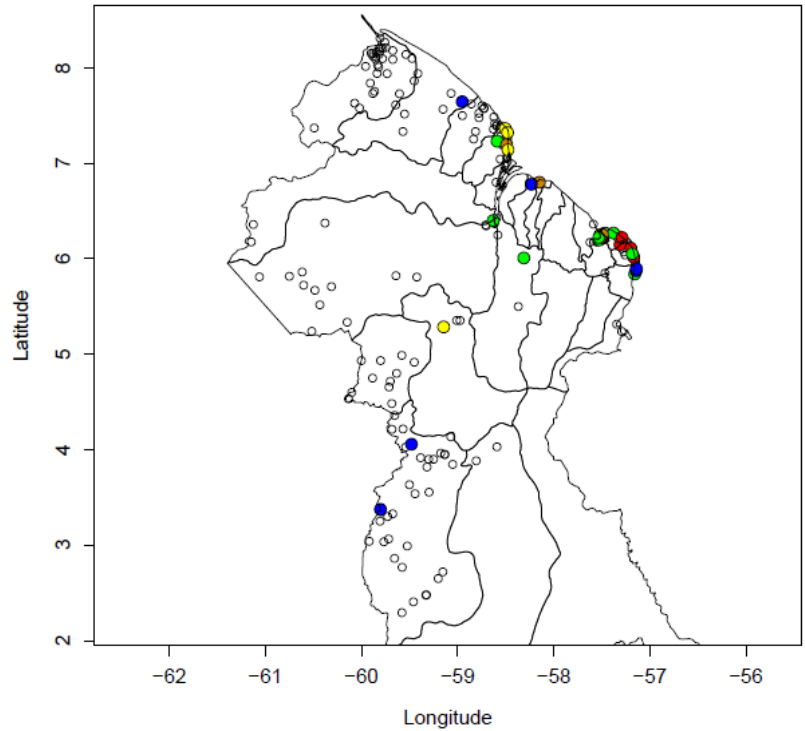

Figure 3. Locations and empirical LF prevalence for 214 villages in Guyana. Locations with zero empirical prevalence are shown as open circles. Locations with positive empirical prevalence colour-coded by quintiles, as follows: 0.009-0.024 (blue); 0.030-0.045 (green); 0.045-0.059 (yellow); 0.059-0.118 (brown); 0.125-0.500 (red)

## Supplementary material

N/A

## Ethics

The Lymphatic filariasis transmission surveys carried out from 2018 to 2021 had their protocols reviewed and approved (#411/2018; #599/2019; #82/2021) by the Guyanese Ministry of Health's Institutional Review Board. Informed consents were obtained from all participants. The IDA Impact Survey (IIS) was approved on 10/09/2022 (#071/2022)

## Data Accessibility

Data and R code are available at www.lancaster.ac.uk/staff/diggle/Guyana_IIS/

## Competing Interests

We have no competing interests.

## Funding statement

This work was supported by the Bill and Melinda Gates Foundation (grant number INV-030046)